\documentstyle[11pt,aaspp4,epsfig]{article}

\newcommand\eq{\begin{equation}}
\newcommand\en{\end{equation}}

\lefthead{.}
\righthead{}

\begin{document}
\title{Constraints on Galaxy Density
Profiles from Strong Gravitational Lensing: \\ The Case of B~1933+503}

\author{ J.D. Cohn, C.S. Kochanek, B. A. McLeod}
\affil{Harvard-Smithsonian Center for Astrophysics, 60 Garden St.,
  Cambridge, MA 02138}
\affil{email: jcohn, ckochanek, bmcleod@cfa.harvard.edu}

\author{ \& C.R. Keeton }
\affil{Steward Observatory, University of Arizona, 933 N. Cherry Ave., Tucson,
AZ 85721}
\affil{email: ckeeton@as.arizona.edu}

\begin{abstract}
We consider a wide range of parametric mass models for B~1933+503,
a ten-image radio lens, and identify shared properties of the
models with the best fits.  The approximate rotation
curves varies by
less than 8.5\% from the average value between the innermost and the
outermost image ($1.5h^{-1}$~kpc to $4.1h^{-1}$~kpc)
for models within $1 \sigma$ of the best fit, and the
radial dependence of the shear
strength and angle also have common behavior for the best models.
The time delay between images 1 and 6, the longest delay between
the radio cores, is $\Delta t =  (10.6^{+2.4}_{-1.1})h^{-1}$~days
($\Omega_0=0.3$, $\lambda_0=0.7$) including all the modeling
uncertainties.  Deeper infrared observations, to more precisely
register the lens galaxy with the radio images and to measure
the properties of the Einstein ring image of the radio source's
host galaxy, would significantly improve the model constraints
and further reduce the uncertainties in the mass
distribution and time delay.
\end{abstract}


\section{Introduction}
Strong gravitational lenses can aid in the determination of two
astrophysical quantities of great interest:  
the density profile of massive galaxies and the value of the Hubble constant.

The density profile of massive galaxies is not known.
Massive early-type and late-type galaxies appear to have mass distributions
consistent with flat rotation curves (total mass $M \propto r$, density
$\rho \propto 1/r^2$) on scales outside their cores and inside their
outer edges.  This is most cleanly shown by spiral galaxy rotation
curves (e.g. Rubin, Ford \& Thonnard \cite{rft78}, \cite{rft80}),
elliptical galaxy X-ray halos (Fabbiano
\cite{fabbiano}), and some gravitational
lenses (e.g., Kochanek \cite{csk1654}).  It is also
supported by modern stellar dynamical studies
of early-type galaxies (e.g., Rix et al. \cite{rixetal}).
The central properties of galaxies are characterized by
density cusps, rather than finite core radii.  In particular,
the central luminosity profiles of massive galaxies generically show density
cusps with $\rho \propto r^{-\gamma}$ and $1 \lesssim \gamma \lesssim 2$
(e.g. Faber et al. \cite{faber}).
Halos in dark matter simulations, which
appear to form with $\gamma\simeq 1-1.5$
cusps (e.g. Navarro, Frenk \& White \cite{nfw}, Klypin et al. \cite{Kly},
Moore et al. \cite{Moo98} ), suggest that
the dark matter cores should have similar properties.
However, it is expected that the dark matter profile will be
significantly modified by the subsequent evolution and interaction with
the baryons (e.g. Blumenthal et al. \cite{blum86},
Dubinski \cite{dubin},
Navarro, Eke \& Frenk \cite{nef96},
Gelato \& Sommer-Larsen \cite{gelato99},
Binney, Gerhard \& Silk \cite{binney00}).
Currently popular scenarios for self-interacting dark matter (SIDM, see
e.g. Spergel \& Steinhardt \cite{SIDM1}; Kochanek \& White \cite{SIDM2})
further complicate
predictions for the central density structure of galaxies.
The resulting central density profile of the dark matter in massive galaxies is
not known.  Even less is known about the angular shapes of galaxy
mass distributions, although spiral galaxies are clearly far rounder
than their luminous disks (see Sackett \cite{sackett} for a review)
and ellipticals can be either flatter
or rounder than their stellar profile (see Buote \& Canizares \cite{buca} and
references therein) and are modestly triaxial (e.g. Franx, Illingworth
\& de Zeeuw \cite{franx}).  See, e.g., Rusin and Tegmark (\cite{rusin})
for a recent summary and more references.

Multiple-image gravitational lenses offer a promising, newer probe
of galaxy mass distributions.  The image geometry usually constrains
the mass inside the critical line of the lens to an accuracy
exceeding that of any other method\footnote{The accuracy is almost
always better than $\lesssim 5\%$ including systematics such as the
cosmological model.  As is true for all mass estimation methods,
there is a global scaling with the Hubble constant (distance),
which for the lenses is simply that the mass is proportional
to $h^{-1}$.}.   Unlike other probes of galaxy mass distributions,
lenses also measure the shape of the gravitational potential
near the critical line accurately.  In most systems, however,
this well measured quantity includes contributions from the
lens galaxy and tidal perturbations (either local to the lens
or along the line of sight) which can be difficult to disentangle
(see e.g. Bar-Kana \cite{barkana}, Keeton, Kochanek \& Seljak \cite{kks}).
In the majority of lenses,
our ability to measure the
radial or angular mass distribution is limited by the
paucity of constraints supplied by the small numbers of
images (2 or 4).  There are, however, many lenses with
either a larger number of discrete images or continuous
arcs or rings formed from the host galaxy or extended
radio emission regions.  These are the systems best used to
probe the details of galaxy mass distributions using
gravitational lensing.

Gravitational lenses are also a promising method for determining
the global value
of the Hubble constant by measuring the light propagation
delays between the images
(Refsdal \cite{refsdal64}).  Although time delays have been measured for six
gravitational lenses, the inferred values for the Hubble constant
are limited by
the systematic uncertainties in the mass models for the systems (see, e.g.
Impey et al. \cite{impey}, Koopmans \& Fassnacht \cite{koop},
Williams \& Saha \cite{williams},
Keeton et al. \cite{keeton2k} for recent examples).
The uncertainties can be reduced by
finding additional constraints on the systems with delays (see Kochanek,
Keeton \& McLeod \cite{kkm}), including external constraints on the models, or
by focusing future time delay studies on lenses with
fewer modeling uncertainties.

Thus, the utility of strong gravitational lensing is maximized in
systems with large numbers of constraints.
The focus of this paper is one such system,
the 10 image lens B~1933+503 (Sykes et al. \cite{sykes},
Marlow et al. \cite{marlow}, Chapman et al. \cite{chapman},
Browne et al. \cite{brow},
Biggs et al. \cite{biggs}, Norbury et al. \cite{norbury}).
B~1933+503 consists of two quadruply imaged
radio sources and one doubly imaged radio source produced by a $z_l=0.76$
lens galaxy.  The source redshift is $z_s = 2.62$ (Biggs et al
\cite{biggs}, Norbury et al. \cite{norbury}).
The images range from unresolved to a pair of
thin arcs (see the schematic in Figure 1).

\begin{figure}[h]
\begin{center}
\leavevmode
\epsfxsize=3.5in \epsfbox{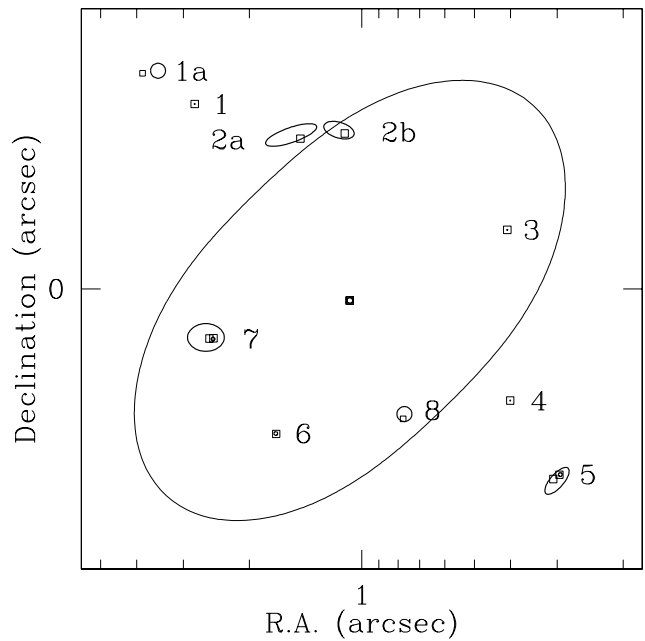}
\end{center}
\caption{ The geometry of B~1933+503.  Images 1,~3,~4~and~6 are four images of
the compact flat spectrum core.  The peaks of images 7 and 5 form a two
image system, while their wings combined with images 2a and 2b form a four
image system.  Finally, images 1a and 8 form a two-image system.  The
positions of the images are shown by their error ellipses, where images
5 and 7 have small uncertainties for their two-image component and large
uncertainties for their four-image component.  The squares show the
image positions found for the best SIE model of the system (see text).
The critical curve for the model is the large ellipse in lighter
type.}

\end{figure}

Nair (\cite{nair}) showed that a simple singular isothermal ellipsoid (SIE)
could fit the available constraints (see also Chapman et al. \cite{chapman}
for some discussion of modeling).  In this paper we extend her analysis
to a wide range of parametric mass distributions to explore the ways in
which the complicated image pattern restricts the mass distribution.
{}From this ensemble of models, we identify common features of the acceptable
models, such as the shape of the mass profile/rotation curve, 
and estimate the uncertainties
in the image time delays for use in determining the Hubble constant.
In \S2 we review the necessary mathematics of lensing and
describe the available data.  In \S3 we catalog the
models we use and their results.  We summarize the results and
suggest further directions in \S4.

\section{Models}

In this section we review the mathematical background and notation for
our models (\S2.1), discuss
the mass distributions we use (\S2.2) and
the available data on B~1933+503 (\S2.3).

\subsection{Notation}

\def\bx{{\bf x}}
\def\bu{{\bf u}}
We consider a thin lens characterized by a projected lens potential
$\phi(\bx)$ which is a function of a Cartesian angular coordinate
$\bx$.  The potential is related to the surface mass density
by $\nabla^2 \phi(\bx) = 2 \kappa(\bx)$ where $\kappa(\bx)$ is
the surface density in units of the critical surface density.  In
addition to the primary lens, we include an external tidal field
defined by an amplitude $\gamma$ and a orientation $\theta_\gamma$.
For a source at angular position $\bu$, the images are found at
solutions $\bx_i$ of the lens equation
\eq
\label{eqone}
 \bu = \bx - \nabla \phi (\bx) - \Gamma \cdot \bx
\en
where the shear tensor is
\eq
\Gamma = \gamma
         \left(
            \begin{array}{rr}
                \cos 2\theta_\gamma & \sin 2\theta_\gamma \\
                \sin 2\theta_\gamma & - \cos 2\theta_\gamma
            \end{array}
         \right).
\en
The shear term models not only local tidal perturbations to the lens
but many of the additional perturbations along the line of sight
(see Kovner \cite{kovner}, Bar-Kana  \cite{barkana}), and it is a necessary
component of any realistic lens model (see Keeton, Kochanek \& Seljak
\cite{kks}).  The inverse magnification tensor of the model is
\eq
 M^{-1} = I - \Gamma -
         \left(
            \begin{array}{rr}
                \phi,_{xx} &\phi,_{xy} \\
                \phi,_{xy} &\phi,_{yy}
            \end{array}
         \right).
\en
and the magnification of an unresolved image is
given by det $M$.  Equation (\ref{eqone}) can also be written
as the condition for an extremum on a time delay surface,
and the surface itself determines the
time delay (see Schneider, Ehlers \& Falco \cite{lensbook}).

The procedures for modeling the positions of lensed point sources are
simple to describe and relatively easy to implement.  Equation (\ref{eqone})
can be read as a map from the image plane to the source plane via
a lens model (potential $\phi$ and external shear $\Gamma$).
The calculations in this paper used
the {\it lensmodel} code (Keeton \cite{lensmodel}). 
For observed image positions $\bx_i$ and model
image positions $\bx_i^{mod}(\bu)$  
the goodness of fit is measured with a simple $\chi^2$ statistic
\begin{equation}
      \chi^2_{point} = \sum_i { |\bx_i -\bx_i^{mod}(\bu)|^2 \over \sigma_i^2 }
\end{equation}
with (in this case) isotropic positional uncertainties $\sigma_i$ for each
image.  The generalization to anisotropic positional uncertainties,
such as will be used here for the arcs, as well as fluxes, is immediate.
The statistic is minimized with respect to the unobserved source position,
$\bu$, and the
parameters of the lens model, where the model parameters may be further
constrained
by measurements of the properties of the lens (position, shape, orientation
$\dots$).
For non-precision work the problem
can be further simplified by simply minimizing the projected separations of the
images on the source plane with various flux, magnification or magnification
tensor
weightings to better mimic the correct image plane statistic (see Kochanek
\cite{csk-mon}).
Most methods using source plane fitting methods are biased to find solutions
with high total magnifications and steep radial density profiles.\footnote{
Roughly writing $\bx_i - \bx_i^{mod} = \delta \bx_i$ for the difference
between model and measured image positions $\bx_i$, and
$\bu - \bu^{mod} = \delta \bu$ one has (from equations 1 and 3)
$\delta \bu = M^{-1}_{ji} \delta \bx_i$.
So if one minimizes $|\delta \bu|^2/\sigma^2$ instead of
$|\delta \bx|^2/\sigma^2$, increasing $M$, the magnification,
will tend to improve the fit.
The corresponding radial profile will tend to be steeper because
steeper profiles generally have higher image magnifications.}

\subsection{Parametrized Models}

We model the mass distribution of the lens
using a wide variety of parametrized ellipsoidal density distributions.
Parametric models of galaxies have the advantages
of being simple and physically motivated.
The disadvantage of parametric models is that they do not span
the complete space of physically realizable mass distributions.
We have chosen a set of parametric models broad enough to
encompass the range of radial mass distributions considered to
represent galaxies.  We have, however, restricted the models
to ellipsoidal surface densities in an external shear field
and thus use a more restrictive range of angular structures
than is in principle possible for a real galaxy.  The various model
families will be discussed in this section and the formulae for their
surface densities can be found in the Appendix.

Before discussing the parameterized models,
it should be noted that
there are non-parametric means of modeling gravitational lenses,
principally the linear programming method of Saha \& Williams (\cite{saha}),
where a gridded, positive definite surface density  distribution is constructed
to fit the data.  Such models avoid the overly restrictive assumptions
of parametric models and allow elegant solution algorithms for the
problem.   Coupled with these advantages are two caveats.
First, the bulk of the mass in a galaxy, particularly the early-type galaxies
that dominate the lens sample, must actually be composed of stars and
dark matter with a positive
definite, quasi-equilibrium {\it phase space distribution function}.
This is more restrictive than a positive definite density distribution, which
is in turn more restrictive than its integral, a positive definite surface
density distribution.\footnote{  In an actual phase space distribution function
there may be some unoccupied orbits, i.e. there can be
some stochasticity due to parts of the galaxy with smaller total
particle numbers such as satellites and globular clusters which may
be responsible for perturbing the image flux ratios in many lenses,
see Mao \& Schneider (\cite{maosch}).}  Many density
distributions permitted
by the Saha \& Williams (\cite{saha}) method are not physically realizable
configurations for galaxies.
Thus, even with symmetry and smoothness constraints, requiring a
positive definite non-parametric real space density allows too much freedom
in the mass model.
Second, the current algorithms solve the lens equations
on the source plane which can introduce the biases
mentioned above.  Nonetheless, as this approach
captures different  features
of the matter distribution than the one considered below, applications
of this method to B~1933+503 would be very interesting.

The parametric models are characterized
by a lens position, $\bx_l$, a mass scale, an axis ratio $q$ (or
ellipticity $e=1-q$), and a major axis position angle, $\theta$.
The major axis density profile is further specified by parameters
describing the radial profile of the galaxy such as core radii,
break radii and exponents.
The models have
surface density distributions $\kappa(m)=\Sigma(m)/\Sigma_c =
\int \, dz \rho(\bx,z)/\Sigma_c$, where $\rho(\bx,z)$ is the density
distribution and $\Sigma_c$ is the critical surface density.
The surface density
$\kappa$ is
a function of the ellipsoidal coordinate
$m^2 = x'^2+y'^2/q^2$,
where $x'$ and $y'$ are Cartesian
coordinates aligned with the major and minor axes of the lens.
The precise parametric forms $\kappa(m)$ of the surface density
distributions for the models are given 
in the Appendix and their properties are discussed below.
In order to simplify the comparison between models below, the
discussion is restricted to radial properties of a spherical density
$\rho(\bx,z) \equiv \rho(r)$.

We start with the singular isothermal ellipsoid (SIE) as a standard model.
It is both analytically convenient (Kassiola \& Kovner \cite{kaskov},
Kormann et al. \cite{kormann}, Keeton \& Kochanek \cite{kkspiral})
and the simplest
mass model which is broadly consistent with our general knowledge of the
mass distributions in observed galaxies as reviewed
above.  The singular isothermal
sphere has a global $\rho \propto 1/r^2$ density profile, which means it
has both a steep central density cusp and a globally flat rotation curve.
We then explore two general families of ellipsoidal density distributions,
varying in radial profile, that contain the SIE as a limit.

The power-law models, whose spherical form is $\rho \propto
(r^2+s^2)^{(\alpha-3)/2}$,
are a simple generalization of the isothermal spheres to include both a
central core radius $s$ and a variable logarithmic slope $\alpha$.  The
isothermal sphere is the limit with $s=0$ and $\alpha=1$, and the power-law
model sequence includes several ``old-fashioned'' models of galaxy luminosity
profiles such as the modified Hubble profile ($\alpha=0$) and the Plummer model
($\alpha=-2$).  This model family was introduced into gravitational lensing
because of its relative analytic simplicity and because observational
evidence at the time suggested that galaxies possessed finite core radii (see
Blandford \& Kochanek 1987; Hinshaw \& Krauss \cite{krauss}).  It has remained
the most popular ``more
complicated than isothermal'' lens model (e.g. Kochanek \cite{csk1654},
Grogin \& Narayan \cite{grogin}, Chae \cite{chae}, Koopmans \& Fassnacht
\cite{koop})
because of its history
and its relative simplicity even as an ellipsoidal model (Barkana
\cite{bar_powlaw}, Chae, Khersonsky \& Turnshek \cite{chae_powlaw}).
Unfortunately,
we now know that galaxies are characterized by central cusp exponents
rather than core radii.
Thus the steeper models, $\alpha<1$ are unphysical while the shallower
models $\alpha>1$ are approximations to
realistic models with shallow density cusps.
We include a sequence of these models with $\alpha <1$ only
for historical continuity.

The pseudo-Jaffe models are related to the Jaffe (\cite{jaffe}) models for
early-type galaxies (see e.g., Keeton \& Kochanek \cite{kkspiral} and
references therein).
They have $\rho \propto (r^2+s^2)^{-1}(r^2+a^2)^{-1}$
for a spherical model, and include both a central core radius $s$ and
an outer break radius $a$.  The model has an
approximately flat rotation curve
for $s \lesssim r \lesssim a$, truncated by the finite central density
inside the core radius and by the finite total mass outside the
break radius.
As we expect the core radius to be zero,
we are mainly interested in the model as a simple probe of
the truncation of the mass distribution at the break radius.  The
pseudo-Jaffe models match the SIE model for $s=0$ and $a\rightarrow\infty$,
and match the $\alpha=-1$ power-law model in the limit $s\rightarrow a$.

We currently lack an implementation of a family of ellipsoidal models
with a variable central cusp exponent (the lensing equivalent of the
$\eta$ models; see Dehnen \cite{dehnen}, Tremaine et al.
\cite{tremaine94}).
The analytic lensing properties of such models were studied by
Evans \& Wilkinson (\cite{ew98}).  In order to
explore the regime of shallow density cusps we
included the NFW (Navarro, Frenk \& White \cite{nfw}), de Vaucouleurs
and $\alpha >1$
models in our survey.  We are interested in the consequences of the
$\rho \propto 1/r$ central density cusp of the NFW models, and not
in its origins as the characteristic density profile of dark matter halos
before modifications due to baryonic matter.
The de Vaucouleurs profile also acts like a shallow, cusped density
distribution ($\rho \propto r^{-5/4}$, see Hernquist \cite{hernquist}), as well
as being the natural
constant mass-to-light ratio model for gravitational lenses.  The NFW models
are characterized by a break radius $a$ between their inner $\rho \propto 1/r$
cusp and their asymptotic $\rho \propto 1/r^3$ profile. The de Vaucouleurs
model effective radius $R_e$ plays a physical role similar to the NFW
break radius.
The $\alpha > 1$ power law models can model the central regions of
galaxies with shallow cusps but cannot be good global mass models as they have
steadily rising rotation curves even at large radii.

\subsection{Data}

We constrain the models using the relative positions of the lens galaxy and
the lensed images and the flux ratios between the images.  Here we summarize
the sources for the constraints and their limitations.  Table 1 presents the
constraints and uncertainties derived from the data in Sykes et al.
(\cite{sykes}),
Nair (\cite{nair}), Marlow et al (\cite{marlow}) and Biggs et al.
(\cite{biggs}).
Figure 1 illustrates the geometry of the system.  The source redshift is
$z_s=2.62$ (Biggs et al. \cite{biggs}, Norbury et al \cite{norbury}) and the
lens redshift is $z_l=0.76$ (Sykes et al \cite{sykes}).  The available
infrared
and optical HST images are too poor to precisely align the radio maps with the
HST images or to perform accurate astrometry or surface photometry of the lens
galaxy (see Marlow et al. \cite{marlow}).

\def\hm{\hphantom{--}}
\def\h0{\hphantom{0}}
\def\mc#1{\multicolumn{1}{c}{#1}}
\begin{deluxetable}{cccccccc}
\tablecaption{Image data}
\tablewidth{0pt}
\tablehead{
ID     &$\Delta$RA     &$\Delta$Dec    &Major         &Minor      &PA of Major
   &Flux     &1 $\sigma$ Error in \\
       &               &               &Axis          &Axis       &Axis 
   &Ratio    &Ratio  }
\startdata
1a     &\hm0\farcs545  &\hm0\farcs584  &0\farcs02\h0  &           & &\hm1.00
&0.44 \\
8      & --0\farcs114  & --0\farcs335  &0\farcs02\h0  &           & & --4.00
&0.44 \\
\hline
1      &\hm0\farcs447  &\hm0\farcs495  &0\farcs001    &           & &\hm1.00
&0.50 \\
3      & --0\farcs389  &\hm0\farcs158  &0\farcs001    &           & & --0.74
&0.37 \\
4      & --0\farcs397  & --0\farcs299  &0\farcs001    &           & &\hm3.50
&1.75 \\
6      &\hm0\farcs230  & --0\farcs387  &0\farcs005    &           & & --0.80
&0.40 \\
\hline
5      & --0\farcs531  & --0\farcs497  &0\farcs005    &           & &\hm 1.00
&0.10 \\
7      &\hm0\farcs398  & --0\farcs134  &0\farcs005    &           & &--1.17
&0.12 \\
\hline
2a     &\hm0\farcs189  &\hm0\farcs412  &0\farcs072    &0\farcs020 &\h0
--71$^\circ$  &         &     \\
2b     &\hm0\farcs061  &\hm0\farcs425  &0\farcs042    &0\farcs021 &
--107$^\circ$  &         &     \\
5      & --0\farcs522  & --0\farcs514  &0\farcs045    &0\farcs018 &\h0
--41$^\circ$  &         &     \\
7      &\hm0\farcs417  & --0\farcs130  &0\farcs049    &0\farcs037 &\h0
90$^\circ$  &         &     \\
\hline
galaxy &\hm0\farcs041  & --0\farcs043  &0\farcs040    &           & &0.00
&0.017 (3 $\sigma$)\\
& & & & & & &(rel. to 1/3/4/6)  \\
\enddata
\tablecomments{\label{imagedata}
The data are derived from Sykes et al. (\cite{sykes}), Nair (\cite{nair}),
Marlow et al. (\cite{marlow}),
and Biggs et al. (\cite{biggs}).  Images 5 and 7 appear twice, once as a
two-image system and once as
a four-images system with images 2a and 2b.  No flux constraints
were used in the latter case.
The coordinates are centered at RA~19$^h$34$^m$31.296$^s$,
Dec~50$^\circ$25$'$22\farcs519 (J2000).
The astrometric uncertainties (Major, Minor and PA) define the error
ellipse, which is circular if
there is only an entry for the Major axis.  The flux ratios have signs
corresponding to their absolute
parities.
The flux ratio assigned to the galaxy is the upper limit on the total flux of
any central
image and corresponds to a flux of $0.072$~mJy in the 8~GHz Biggs et al.
(\cite{biggs}) maps.
}
\end{deluxetable}

The lensed images (see Figure 1) are composed of three sets of multiply imaged
source
components.  There are four images of the flat spectrum core (image 1, 3, 4 and
6)
which have very precise astrometry from the Sykes et al. (\cite{sykes}) VLBI
observations.  There are two images of one steeper spectrum component
(images 1a and 8) whose astrometry is relatively poor because of their low
fluxes
and proximity to the brighter core images.  There are four images of a second
steeper spectrum component (images 2a, 2b, 5 and 7) but these cannot be naively
treated as four images of the same source point because of their extended
structure (see Nair \cite{nair}).  In particular, the peaks of images 5 and
7 are only doubly imaged and do not have counterparts in images 2a and 2b.
There is no gap in the arc formed by images 2a and 2b, so they contain only
partial images of the source component producing images 5 and 7.  Following
Nair (\cite{nair}) we treat the peaks of images 5 and 7 as a two-image
system.  We then consider all four images 2a/2b/5/7 as a four-image system,
using error ellipses corresponding to the major and minor axes of the images
to compensate for our uncertainty in precisely how the extended structure
should be modeled.

The flux ratios of the images depend on both frequency and time.  To derive
the flux ratios of the images we averaged the flux ratio measurements by
Sykes et al. (\cite{sykes}) and Biggs et al. (\cite{biggs}). The differences
are probably dominated by systematic differences in the resolution and
sampling,
temporal variability, and intrinsic systematic problems with flux ratios
(see Mao \& Schneider \cite{maosch}) rather than simple measurement errors.
Thus, the problem with flux ratios is always the precision with which they
should be imposed.  For images 1,~3,~4~and~6, we used 50\% errors on the fluxes
because the large flux ratio between image 4 and the other images is unlikely
to be
due to the effects of lensing -- generic lens configurations do not generate
such flux ratio patterns for point images.  For images 1a and 8 we used the
reported flux errors. For images 5 and 7 modeled as a two-image system,
we used 10\% flux errors (which we view as a lower bound on any flux ratio
constraints), and for images 2a,~2b,~5~and~7 modeled as a four-image system
we used no flux constraints because we are modeling extended rather than
point images.  We will explore the consequences of changing these estimates
of the uncertainties on the results.

No central radio image is detected in any of the
radio maps, which is a powerful
constraint on the central surface density of the lens.
In our standard models
we used a detection limit for central images of
0.072~mJy based on the 3$\sigma$
detection limit in the 8~GHz observations by Biggs et al. (\cite{biggs}).
This is very conservative, both because we
use a 3$\sigma$ limit as a 1$\sigma$
limit and because we apply it to the central
image for each image system separately
rather than to the sum (i.e. if a model produced an
0.072~mJy central image in its
model for each of the 1a/8, 1/3/4/6 and 5/7
image systems we will count it as a
3$\sigma$ detection instead of a 9$\sigma$ detection).   We will explore the
consequences of using a tighter 0.024~mJy (1$\sigma$) constraint.
No position constraint was imposed on this undetected central image.

\begin{figure}[h]
\begin{center}
\leavevmode
\epsfxsize=5in \epsfbox{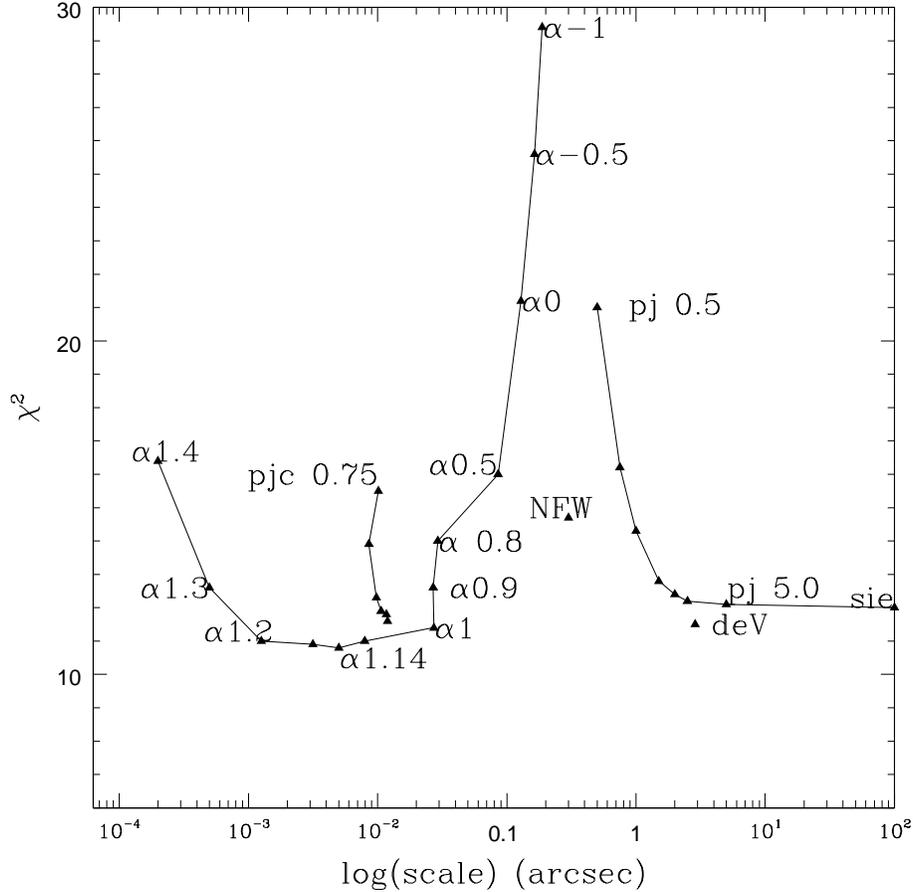}
\end{center}
\caption{
The $\chi^2$ values for the best fit models in each class ordered by their
scale radii.  The power law models (the sequence labeled
$\alpha n$ for $\alpha=n$) are
shown at the best fit core radius $s$ for $\alpha \leq 1$.  For $\alpha >1$
where the core radius is unimportant the models are simply offset horizontally
to distinguish the points.  The $\chi^2$ of the pseudo-Jaffe models
are shown as a function of the outer break radius $a$ (the sequence 
labeled pj$a$ for break radius $a$).  The reference SIE model
is shown at $\log(scale)=2$ because it is the limit of the pseudo-Jaffe
sequence as $a\rightarrow \infty$.  The SIE model is almost identical
to the $\alpha=1$ softened power law model.  The best NFW and de Vaucouleurs
(labeled deV) models are shown at the appropriate NFW break radius $\log(a)$
and de Vaucouleurs effective radius $\log(R_e)$.}
\end{figure}

\section{Results}

We now describe the results of modeling the system with the density
distributions
discussed in \S2.1.  These consists of three discrete models (SIE, NFW and deV)
and two two-parameter model families (the power law models and the pseudo-Jaffe
models).  All the models use ellipsoidal density distributions and include an
independent external shear.  Models without the shear terms fit the data so
poorly
that we do not discuss their properties.  The necessity of an independent
external
shear is expected (see Keeton, Kochanek \& Seljak \cite{kks}).
The external shear required varied from between 2\% and
14 \% for the best 
($2 \sigma$ models), and the galaxy positions varied by
$\le 10$ mas.  The images will
strongly constrain the mass distribution over the region containing images
(from $1.5h^{-1}$~kpc to $4.1h^{-1}$~kpc) with the mass being most tightly
constrained
near the critical radius of $2.6h^{-1}$~kpc, where  $1\farcs0=5.1h^{-1}$~kpc
(for $\Omega_0=0.3$, $\lambda_0=0.7$, and $H_0=100h$~km~s$^{-1}$~Mpc$^{-1}$).
Table 2 summarizes the results.

\subsection{Statistical Summary}

The data supply a total of 26 constraints, and the models have $N_p=7$ or
8
parameters, leaving us with $N_{dof}=19$ or $18$ degrees of freedom.  We will
explore the results using standard results for Gaussian statistics, which we
now summarize.
\begin{enumerate}
\item Comparing different models:  For $N_{dof}$ degrees of freedom, the value
of our
fit statistic should follow a $\chi^2$ distribution with $N_{dof}$ degrees of
freedom
with a mean value of $N_{dof}$ and a variance of $(2N_{dof})^{1/2}$.  For
large $N_{dof}$ the distribution is asymptotically Gaussian, a limit we can use
relatively
safely.  Thus, for $N_{dof}=18$ the mean and variance of the $\chi^2$
distribution
are $18$ and $6$ respectively.

\item Rescaling the statistics: The best
models have $\chi^2 \simeq 11$ for $N_{dof}=18$
which is somewhat low although statistically
possible (23\% of models with $N_{dof}=18$
would have $\chi^2 < 11$ or $>25$ for Gaussian statistics).
However, we may have
overestimated the uncertainties or degrees of freedom
in the model.  We can adjust
for this by rescaling the models to make the best fit
model have $\chi^2_{min}=N_{dof}$,
which corresponds to shrinking the typical constraint
uncertainties by 20\% (an entirely
plausible correction).  Since we do not rescale the fit statistics, we may be
overestimating the range of acceptable models and the uncertainties on their
parameters.

\item Parameter significance: For a multi-parameter model
like the power-law models,
the introduction of a parameter (e.g. adding a core
radius to the SIE to fit the
more general isothermal ellipsoid) is statistically
significant if the change in
the fit statistic with the addition of the
parameter, $|\Delta\chi^2|> f_{lim} \chi^2/N_{dof}$,
exceeds the results without that parameter by a factor $f_{lim}=0.5$,
$3.0$ or $8.1$ for significance levels of 50\%, 90\% and 99\% compared to the
improvement expected for a random variable.  These limits are based on the
F-test
for adding one variable to a model with $N_{dof}=18$ or $19$
degrees of freedom.

\item Parameter estimation: For a multi-parameter model, parameter ranges are
estimated from the change in the fit statistics relative to the best model,
$\Delta\chi^2= \chi^2-\chi^2_{min}$ where the parameter range for one variable
is acceptable at the 1$\sigma$, 90\%, 2$\sigma$ and 99\% confidence level
for  $\Delta\chi^2=1.0$, $2.71$, $4.00$ and $6.63$ respectively.
Upper bounds and parameter ranges quoted below will be $1 \sigma$ bounds
($\Delta \chi^2 = 1$) unless otherwise noted.
\end{enumerate}

\begin{deluxetable}{clccrrll}
\tablecaption{The Statistical Properties of The Fits}
\tablewidth{0pt}
\tablehead{
  Family  &\mc{Type} &$N_p$ &$N_{dof}$  &$\chi^2$ &\mc{$s$} &\mc{$a$ or $R_e$}
&Comments }
\startdata
\hline
  SIE         &                  &7 &19   &12.0  &--                   &-- \\
  NFW         &                  &8 &18   &14.7  &--
&$a=0\farcs300$ \\
  deV         &                  &8 &18   &11.4  &--
&$R_e=2\farcs86$ \\
\hline
power-law  &$\alpha= \phantom{-}1.4$  &7 &19  &16.4 &$\equiv$0\farcs0001 &-- \\
        &$\alpha= \phantom{-}1.3$  &7 &19   &12.6  &$\equiv$0\farcs0001  &-- \\
       &$\alpha= \phantom{-}1.2$  &7 &19   &11.0  &$\equiv$0\farcs0001  &-- \\
 &$\alpha= \phantom{-}1.14$  &7 &19   &10.8  &$\equiv$0\farcs0001  &-- \\
       &$\alpha= \phantom{-}1.1$  &7 &19   &11.0  &$\equiv$0\farcs0001           &-- \\
         &$\alpha= \phantom{-}1.0$  &8 &18   &11.4  &0\farcs027           &--
&isothermal ellipsoid \\
         &$\alpha= \phantom{-}0.9$  &8 &18   &12.6  &0\farcs027           &--
& \\
   &$\alpha= \phantom{-}0.8$  &8 &18   &14.0  &0\farcs029           &--
& \\
   &$\alpha= \phantom{-}0.5$  &8 &18   &16.0  &0\farcs086           &-- \\
        &$\alpha= \phantom{-}0.0$  &8 &18   &21.2  &0\farcs128           &--
&modified Hubble      \\
              &$\alpha=   -0.5$  &8 &18   &25.6  &0\farcs164           &-- \\
              &$\alpha=   -1.0$  &8 &18   &29.4  &0\farcs188           &-- \\
             \hline
pseudo-Jaffe  &$s\equiv0$        &7 &19   &12.1  & --
&$a\equiv5\farcs0$ \\
              &$s\equiv0$        &7 &19   &12.4  & --
&$a\equiv2\farcs0$ \\
              &$s\equiv0$        &7 &19   &21.0  & --
&$a\equiv0\farcs5$ \\
              &$s\ne 0$          &8 &18   &11.6  &0\farcs012
&$a\equiv5\farcs0$ \\
\enddata
\tablecomments{\label{statistics}
The model families and parameters are
described in \S2.2 and Appendix A. Figures 2 and 3
compare the models graphically.  The number of parameters
varied is $N_p$, leading to
a model with $N_{dof}$ degrees of freedom and
goodness of fit $\chi^2$.   We only
present representative pseudo-Jaffe models as the
best fit model corresponds to the
softened isothermal ellipsoid (the limit $a\rightarrow\infty$).
The SIE is the
same as the softened isothermal model ($\alpha=1$) in the
limit $s\rightarrow 0$,
and the $\alpha=-1$ power law model is the same as the
pseudo-Jaffe model in the
limit that $s\rightarrow a$.  }
\end{deluxetable}

\subsection{Model Properties}

We start with the standard SIE model
with external shear, which has $\chi^2/N_{dof}=12.0/19$.  The model
results are compared to the constraints in Figure 1 and are remarkably good,
given the incredible simplicity of the model.  The SIE model
is a particular example
of both the power law and pseudo-Jaffe two-parameter models,
which we consider next.
Adding a core radius to the SIE, making a general
isothermal ellipsoid, slightly
improves the fit to  $\chi^2/N_{dof}=11.4/18$ but the improvement is not
statistically
significant (66\% for the F-test).
The formal core radius is
$\sim 140 h^{-1}$~pc, but the more useful result
is an upper limit of $\sim 260 h^{-1}$~pc.
The fit is not improved by adding an outer break radius $a$ to the isothermal
density profile (the pseudo-Jaffe models).
The value of $\chi^2$ monotonically
rises as the break radius is reduced from the
SIE limit ($a\rightarrow \infty$),
leading to a lower bound of $ a > 1\farcs4 \simeq 7.1 h^{-1}$~kpc.
The core
radius of the pseudo-Jaffe model plays the same role as that of the isothermal
ellipsoid -- a small core radius slightly, but insignificantly, improves
the fit.

The traditionally favored power-law models, with $\alpha < 1$, a finite core
radius, and a more rapidly declining asymptotic density distribution than the
isothermal models, have increasingly large values for $\chi^2$ as the mass
distribution becomes more centrally concentrated (see Figure 2).  These models
require large core radii to fit the data (e.g. 
$(653^{+ 92}_{-97})h^{-1}$~pc for the modified
Hubble profile, see Figure 3).  In addition, for
$\alpha < 0$ models, external shears of $>20\%$ are required,
increasing as $\alpha$ decreases.  These large external shears
make the interpretation of the parameters questionable--consequently 
we have cut off our analysis at the $\alpha =-1$
model (shear $\sim 23 \%$); the $\alpha < -1$ models have
$\chi^2 >35$ and external shears of $>27\%$.
The power law models are consequently unacceptable
unless nearly isothermal ($\alpha \simeq 1$), as we might expect from
our current understanding of the structure of galaxies (see \S1).  The
poor fit of the models is not solely due to the restriction on the
visibility of the central image, as these models remain unacceptable
when we weaken or eliminate the constraints on the image
flux ratios (see below).

The fit does improve significantly for power-law
models with $\alpha > 1$, where the
models are mimicking galaxies with softer central density cusps
than the SIE model.
For these models the core radius
must be small, and for simplicity we only show the results with the core
radius fixed.\footnote{For example, the $\alpha=1.1$ power-law model
has the same $\chi^2$ for an optimized core radius
 ($0\farcs0002$)
and for the core radius fixed as in Table 2.}
The best fit is found for $\alpha = 1.14^{+0.11}_{-0.17}$ 
which corresponds to
a density cusp with $\rho \propto r^{-1.86^{+0.11}_{-0.17}}$
The exponent $\alpha$,
unlike the core radius $s$, is a statistically significant new parameter
(82\% for the F-test).
Unfortunately, the $\alpha >1$ models are poor
global models because of their steadily rising rotation curves.

The NFW and de Vaucouleurs (deV) models provide examples of models with
shallow central cusps (near $\rho \propto 1/r$) and are reasonable global
models for the mass distribution.  (The $\alpha>1$ models cannot produce
such shallow cusps because as $\alpha \rightarrow 2$ they become constant
surface density sheets rather than $\rho\propto1/r$ cusps because of how
the limits are ordered.)  Figure 3 shows the variation in $\chi^2$ with
the de Vaucouleurs effective radius $R_e$, the NFW break radius $a$,
and, for comparison, the core radius $s$ of the isothermal ellipsoid
($\alpha=1$) and the pseudo-Jaffe model break radius $a$. The NFW
model is barely acceptable at 2$\sigma$ significance, with
a break radius of $a=(1.53^{+0.36}_{-0.26})h^{-1}$
~kpc.  The more cuspy de Vaucouleurs models
produce acceptable fits for an effective radius of $R_e=(14^{+9}_{-4})h^{-1}$
~kpc.
This is much larger than is realistic for a constant mass-to-light ratio model
of the lens galaxy.\footnote{While the HST images are not adequate to fit a
photometric model, the optical and IR emission
is confined to a region comparable
in size to the lens, so a real constant mass-to-light ratio model would have
$R_e < 1\farcs0$ ($5h^{-1}$~kpc).}

\begin{figure}[h]
\begin{center}
\leavevmode
\epsfxsize=3.5in \epsfbox{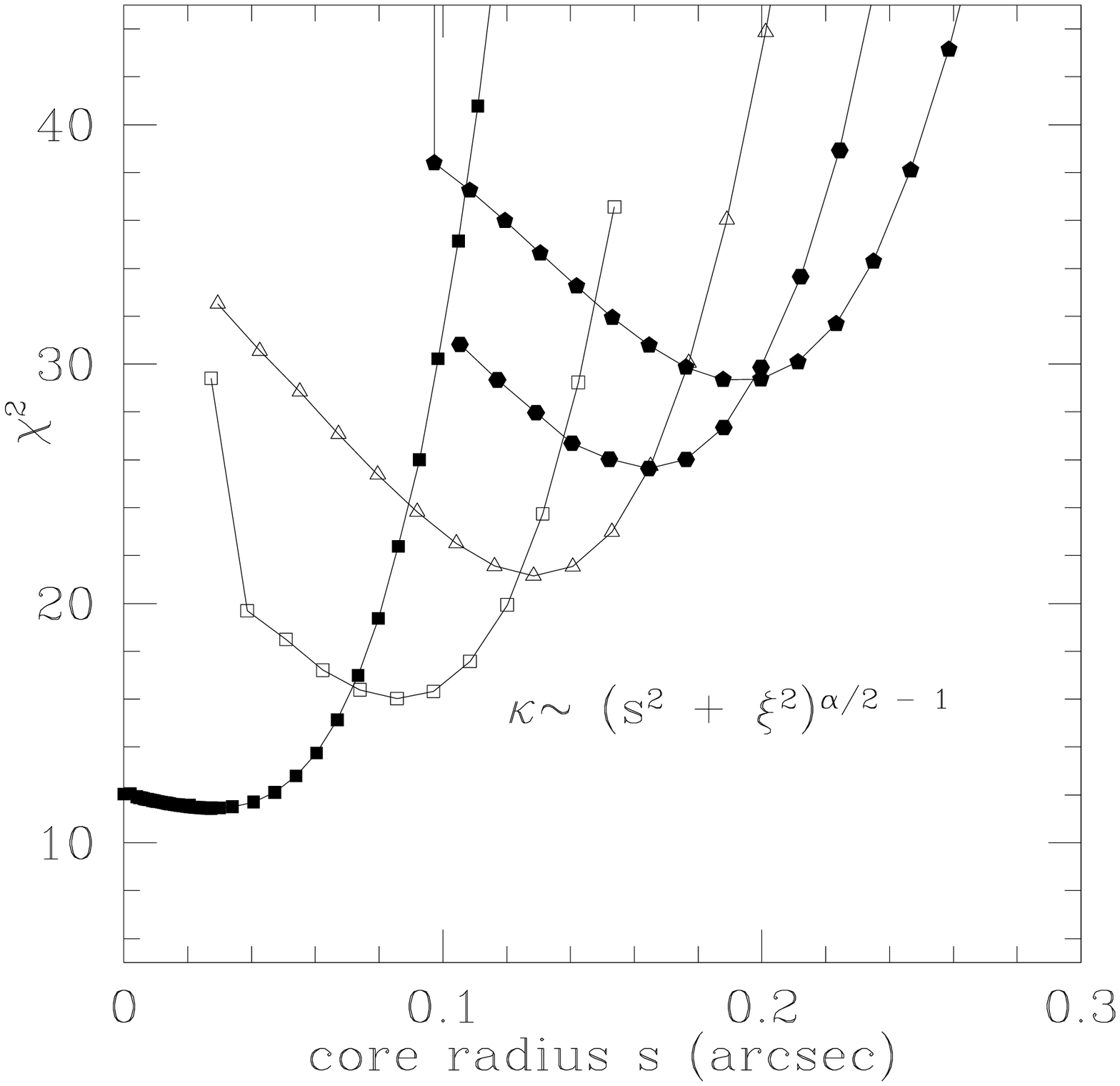}
\epsfxsize=3.5in \epsfbox{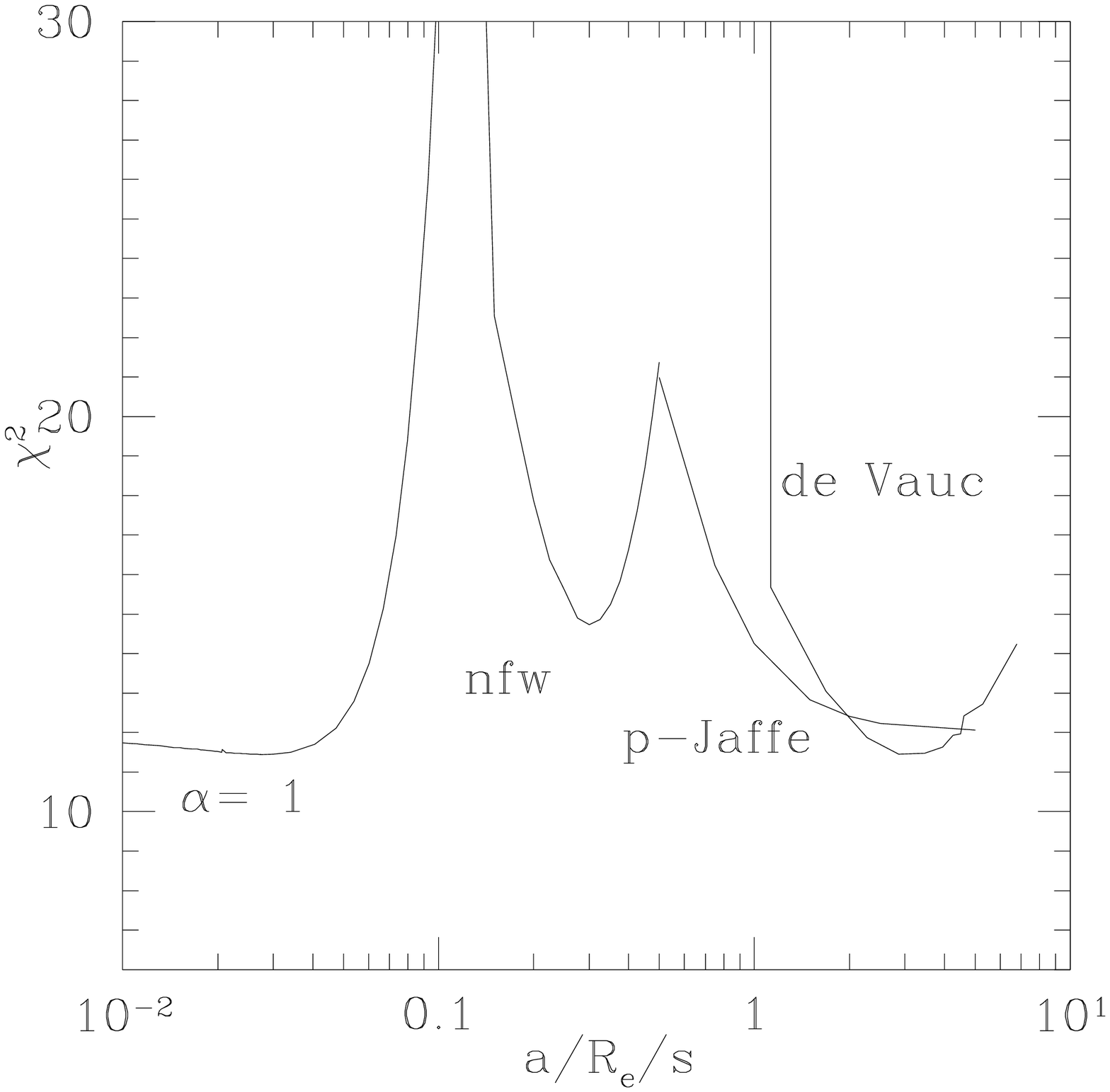}
\end{center}
\caption{
(Left) The $\chi^2$ for the power law models as a function of the core radius
$s$.  Results are shown for
fixed exponents of $\alpha = 1.0$
(the softened isothermal model), $0.5$, $0.0$ (the
modified Hubble model), $-0.5$, $-1.0$
going from left to right at the minima.
(Right) The $\chi^2$ as a function of the
NFW break radius $a$ and
the de Vaucouleurs effective radius $R_e$
as compared to the softened
isothermal core radius $s$ and the
pseudo-Jaffe outer break radius
$a$.
}
\end{figure}

The uncertainties we used for the image flux ratios are somewhat
arbitrary (see \S2.3),
so we explored the effects of changing these limits. 
We recalculated all the models with 50\% flux
uncertainties
for all images ($\chi^2_{min}/N_{dof}=9.0/19$), no flux constraints at all
($\chi^2_{min}/N_{dof}=6.7/10$), and using the larger of the quoted errors in
the flux measurements or 10\% of the flux 
($\chi^2_{min}/N_{dof}=61.5/19$).
 In
all three of these cases the best-fit model was the SIE, or equivalently the
$\alpha=1$ isothermal ellipsoid with the statistically insignificant
addition of a 
core radius.  
The $\alpha=1.14$ best-fit model for our standard 
flux ratio uncertainties remains within 1$\sigma$ of the best fitting models
for two cases, and had a $\Delta \chi^2 = 1.4$ for the case where
10\% flux limits are used.
In the case where no flux limits are imposed the $\alpha < 1$ models are
still disfavored, with $\alpha > 0.5 (1 \sigma), \; -0.4 (2 \sigma)$.
The overall statistical ordering of the models
is essentially unaffected by changes in the choice for the flux ratio
uncertainties.
We also explored effects of changing the limits
on the flux
of any central images, discussed below. 

\begin{figure}[h]
\begin{center}
\leavevmode
\epsfxsize=3.5in \epsfbox{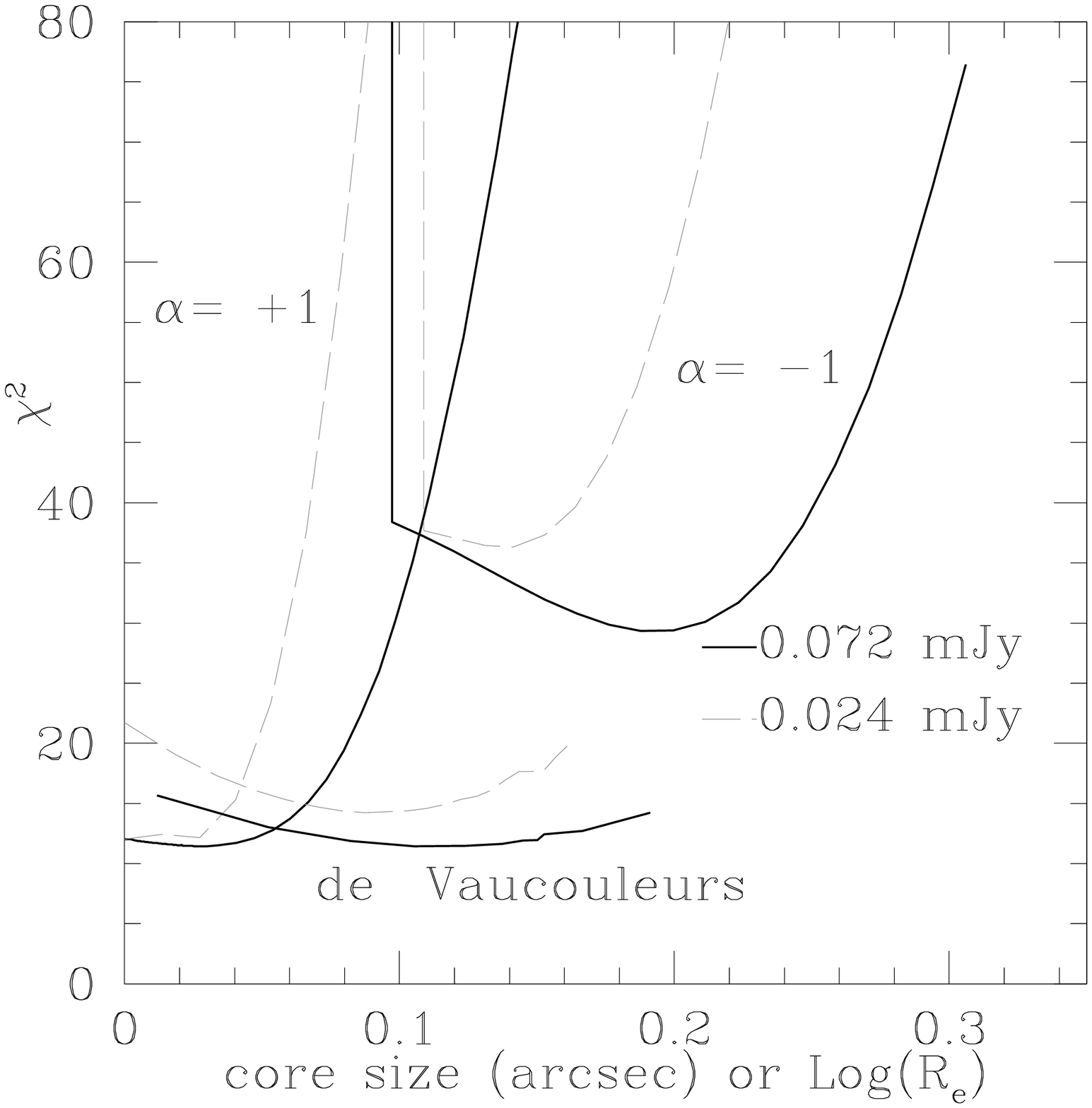}
\end{center}
\caption{The effect of stronger limits on the flux of any central images.
We show
the $\chi^2$ for the softened isothermal model, the
$\alpha=-1$ power-law model
and the de Vaucouleurs model
using either our standard,
weak (3$\sigma$, $0.072$~mJy) constraint (solid lines) or
a stronger
(1$\sigma$, $0.024$~mJy) constraint (dashed lines) on the flux of any central
images.  For model requiring finite
core radii the stronger constraint forces the
$\chi^2$ to increase and the best fit core radius to decrease.
}
\end{figure}

The central cusp exponent or core radius of the
distribution determines the visibility
of the central or ``odd'' image
(Narasimha, Subramanian \& Chitre \cite{nara86}), whose
flux decreases as the central
surface density or concentration of the lens increases.
Thus, the limits on the core
radius depend on the treatment of the central image.
Figure 4 illustrates the consequences of tightening the limit on the flux of
central images from 0.072~mJy (3$\sigma$
in the 8~GHz map) to 0.024~mJy (1$\sigma$)
for the isothermal ellipsoid ($\alpha=1$), the $\alpha=-1$ power-law, and the
de Vaucouleurs models.  The scale radii are forced to decrease, and for the
$\alpha=-1$ power-law model and the de Vaucouleurs model
the $\chi^2$ statistics
rise significantly. The change in $\chi^2$ for models with low $\chi^2$ 
is generally small ($\Delta \chi^2 < 1$).
Mao, Witt \& Koopmans (2000) explored the effects of
adding a central black hole to the lens galaxy as an alternate means of
suppressing the flux of central images.  While we do not expect
the addition of a central black hole to modify the global trends 
discussed below (as model
comparisons were essentially unchanged when the flux constraints were
completely removed), it would be interesting to include central black
holes in future analysis.

Finally, we explored the effects of further generalizing the models.
In particular
we were concerned that the limitation to ellipsoidal models might force an
undesirable correlation between the monopole and quadrupole structures of the
lens galaxy.  We tested this by modeling the galaxy as a combination of two
pseudo-Jaffe models where the inner cutoff radius of one equaled the outer core
radius of the other (see Keeton et al. \cite{keeton2k})
and then optimized the relative masses, ellipticities and orientations of
the two components.  The best fit models had $\chi^2 \sim 10-11$ and 2 or 3
fewer degrees of freedom, so the improvements in the fit were not statistically
significant given the reduction in the number of degrees of freedom.
Modeling this system as a combination of two other models
(Keeton \cite{keeton00}) produced similar behavior in the monopole and
quadrupole structures discussed below.

\subsection{Global trends}
Parametric models make it difficult to see the global trends implied by the
range of acceptable models or to compare the physical properties of the
individual models.  We can illustrate the similarities and differences
of the models by looking at the monopole and quadrupole deflection profiles
of the models (see e.g. Kochanek \cite{csk-mon} for definitions).  
For a potential such that $\nabla^2\phi=2 \kappa(r,\theta) = 2
\Sigma(r,\theta)/\Sigma_c$,
the monopole deflection of the lens is
\begin{equation}
\label{monopole}
  \alpha_0(r) = { 1 \over r } \int_0^r r' dr' \int_0^{2\pi} d \theta \nabla^2
\phi(r)
              = { 2 \over r } \int_0^r r' dr' \int_0^{2\pi} d \theta \frac
{\Sigma(r,\theta)}{\Sigma_c} = {2 M(<r) \over {\Sigma_c r} }
\end{equation}
where $M(<r)$ is the mass enclosed by the circle and the
coordinates are centered on the lens galaxy.
The rotation curve of the galaxy is roughly proportional
to the square root of the deflection monopole (projection effects
modify this slightly).
Generic deflection or rotation curve
profiles have three radial regimes. At small radii
there is a core region where the
deflection can go to zero either because of a finite
core radius or the presence of a
shallow density cusp.  The size of the region is set either by the
core radius or the
break radius where the steep outer density cusp is matched to the
shallower inner
cusp.  In the inner region the rotation curve rises, and can become
flat for a while.
At large radius the distribution
must have a break beyond which the rotation curve becomes Keplerian (or near
Keplerian for the NFW model).

Figure 5 shows the monopole as a function
of radius for the models from Table 2.
As is typical of lens models (see Kochanek \cite{csk-mon}),
all models agree very
precisely on the mass enclosed by the mean critical line of the lens.  For a
critical surface density of  $\Sigma_c =0.62h$~g~cm$^{-2}$ ($\Omega_0=0.3$,
$\lambda_0=0.7$),
the mass inside $r=0\farcs5$ is $6.0 \times 10^{11} h^{-1} M_\odot$ independent
of the
parametric model.  The statistically acceptable models have slightly rising
or falling deflection profiles, and since $\alpha_0 \propto v_c^2$ the
rotation curve of the acceptable models varies by $<$ 8.5\% from the
average
between the
inner and outer images ($1.7h^{-1}$~kpc to $4.1h^{-1}$~kpc).
The model with the least flat rotation curve ($\alpha=1.3$)
has a core of less than $0\farcs0001$, and so projection effects should not
alter this limit significantly (for the spherically symmetric case the
change is less than $0.01\%$).
The models with
a slightly falling rotation curve are the pseudo-Jaffe models, and the models
with the rising rotation curves are the $\alpha >1$ shallow-cusp power law
models.  The worst fitting models from Table 1 tend to have rapidly rising
and then falling rotation curves as shown on the right in figure 5.

\begin{figure}[h]
\begin{center}
\leavevmode
\epsfxsize=3.5in \epsfbox{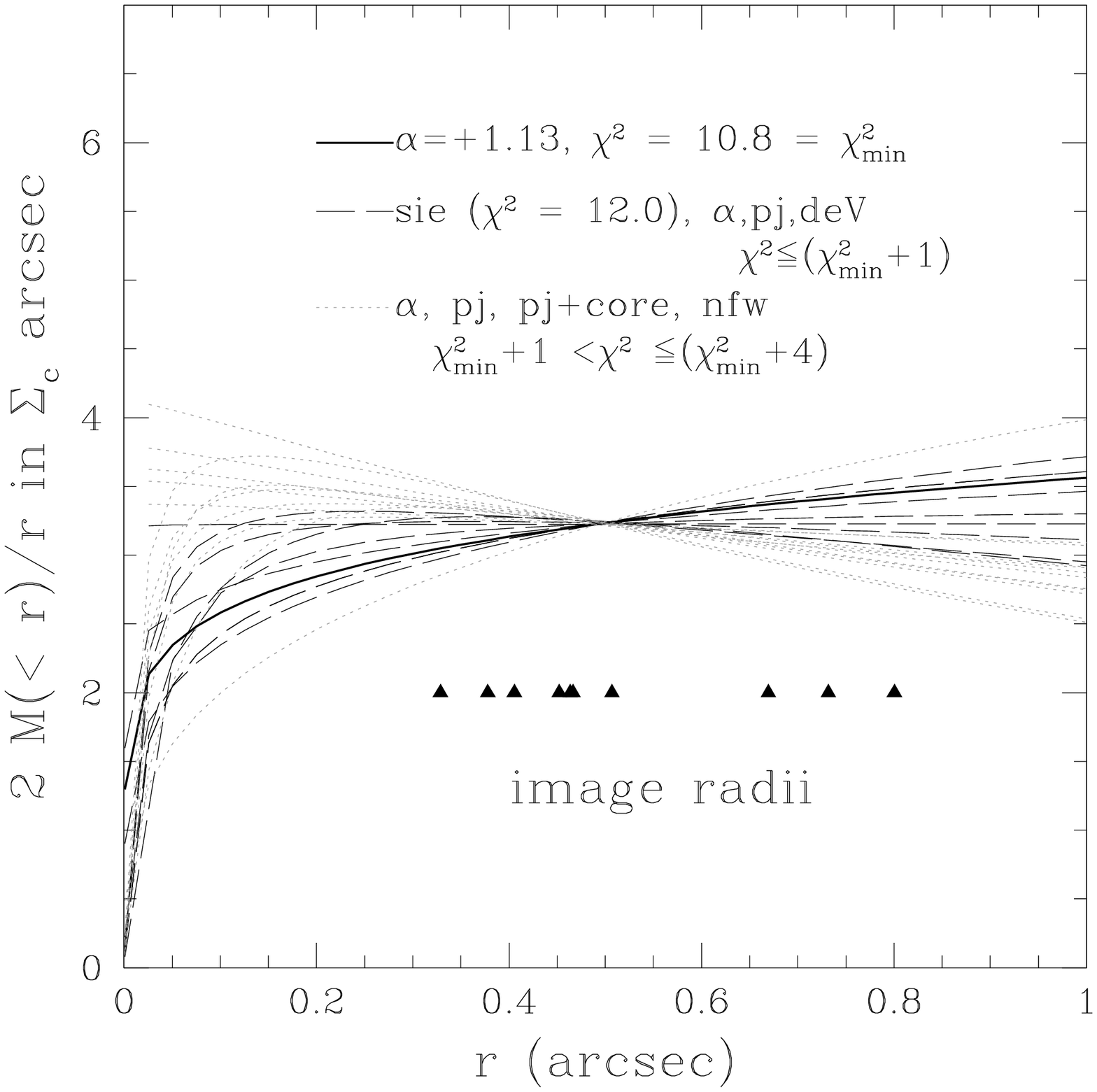}
\epsfxsize=3.5in \epsfbox{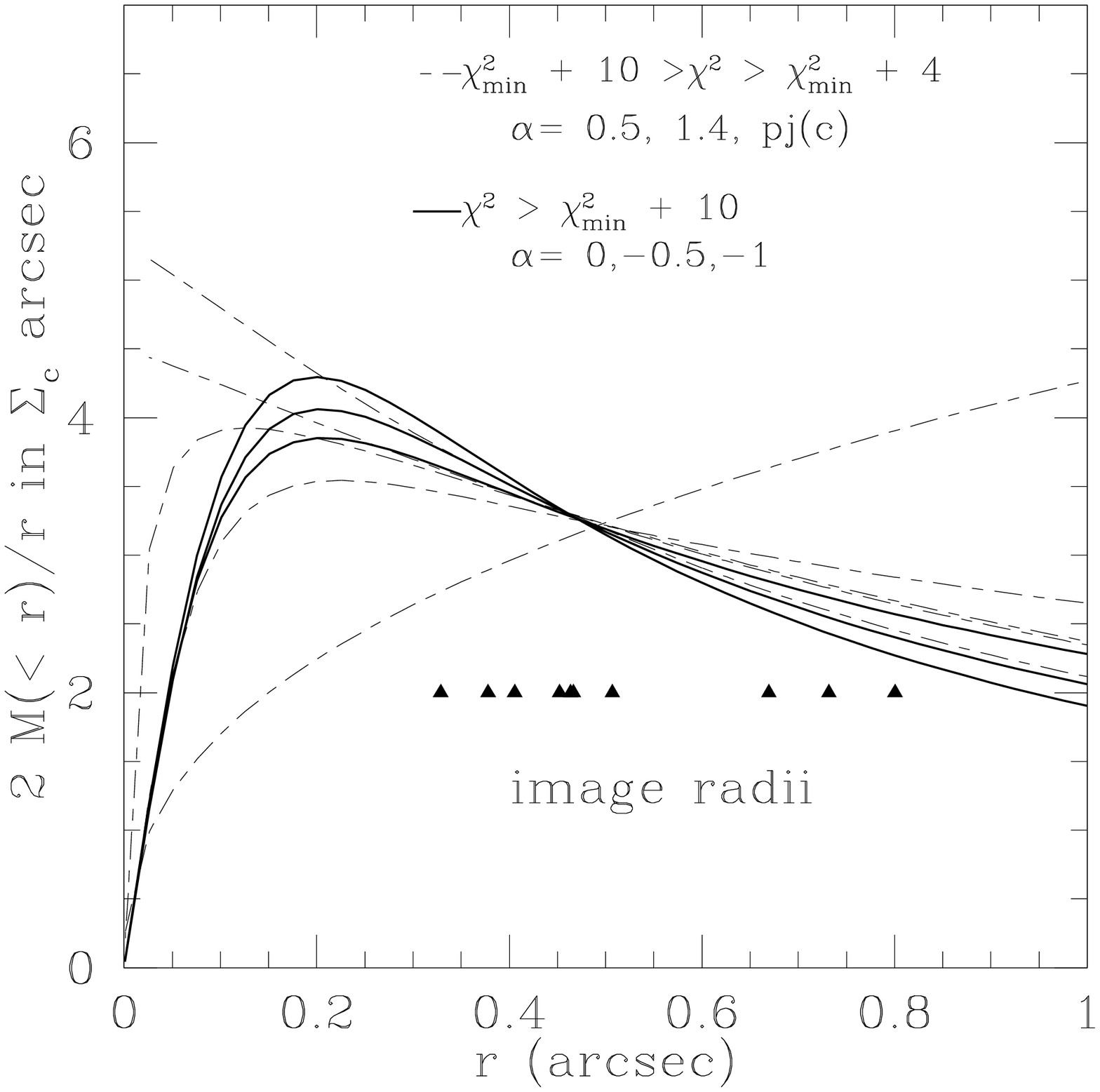}
\end{center}
\caption{The monopole deflection of
acceptable (left) and unacceptable (right) mass
models.  The monopole deflection is proportional
to $v_c^2 \simeq M(<r)/r$ where $M(<r)$ is the
mass enclosed by a circular aperture of radius $r$.  The circular velocity is
roughly proportional to the square root of the deflection, $(M(<r)/r)^{1/2}$.
The models with acceptable fits have $\chi^2 < \chi_{min}^2 + 4$
and the models with
unacceptable fits have $\chi^2 > \chi_{min}^2 +4$.
The line types distinguish between
models in different $\chi^2$ ranges.
The triangles below the curve show the radial positions
of the 10 images.  The constraints on the mass distribution are strongest
in the annulus containing the images.
}
\end{figure}

\begin{figure}[h]
\begin{center}
\leavevmode
\epsfxsize=3.5in \epsfbox{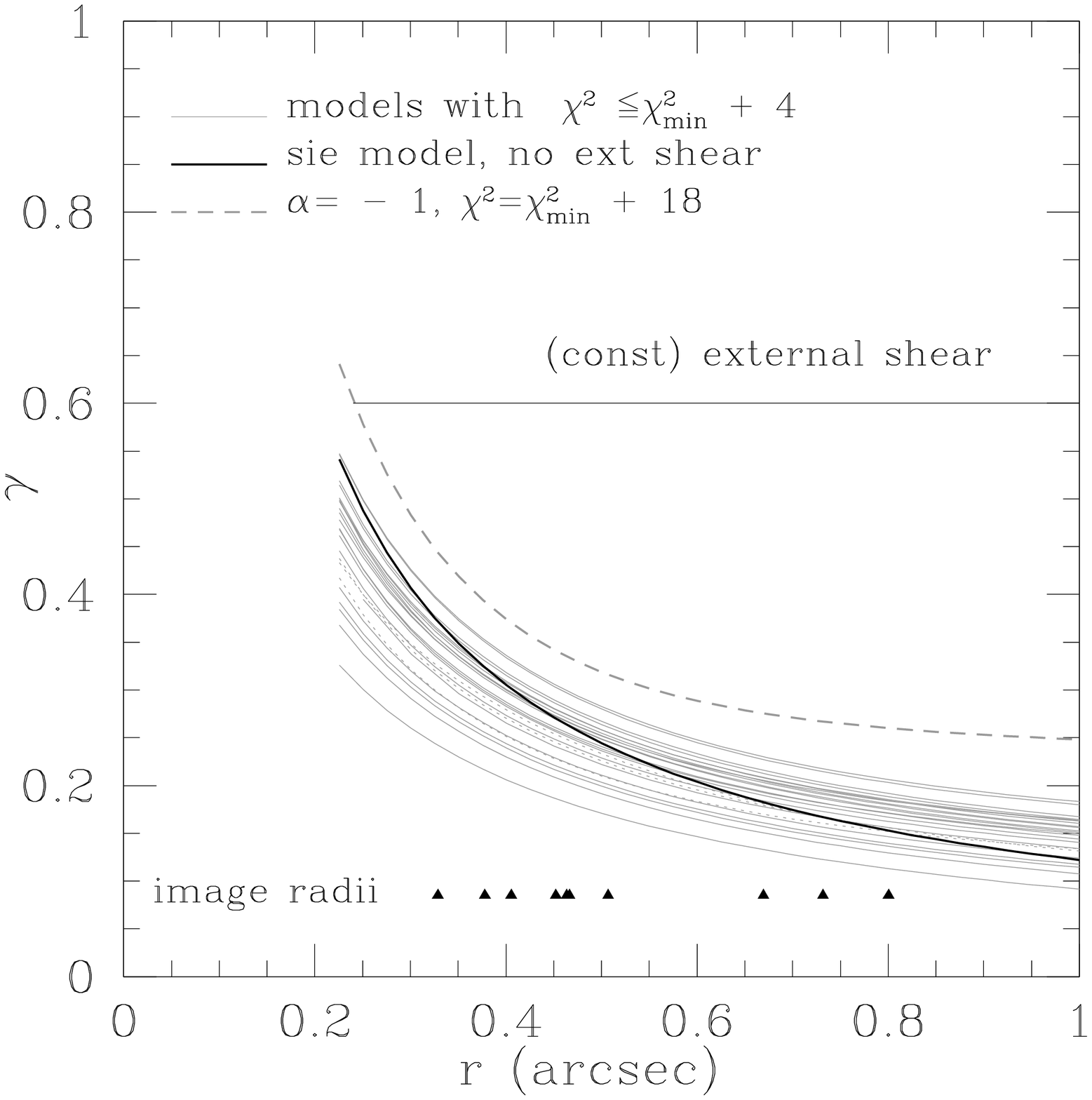}
\epsfxsize=3.5in \epsfbox{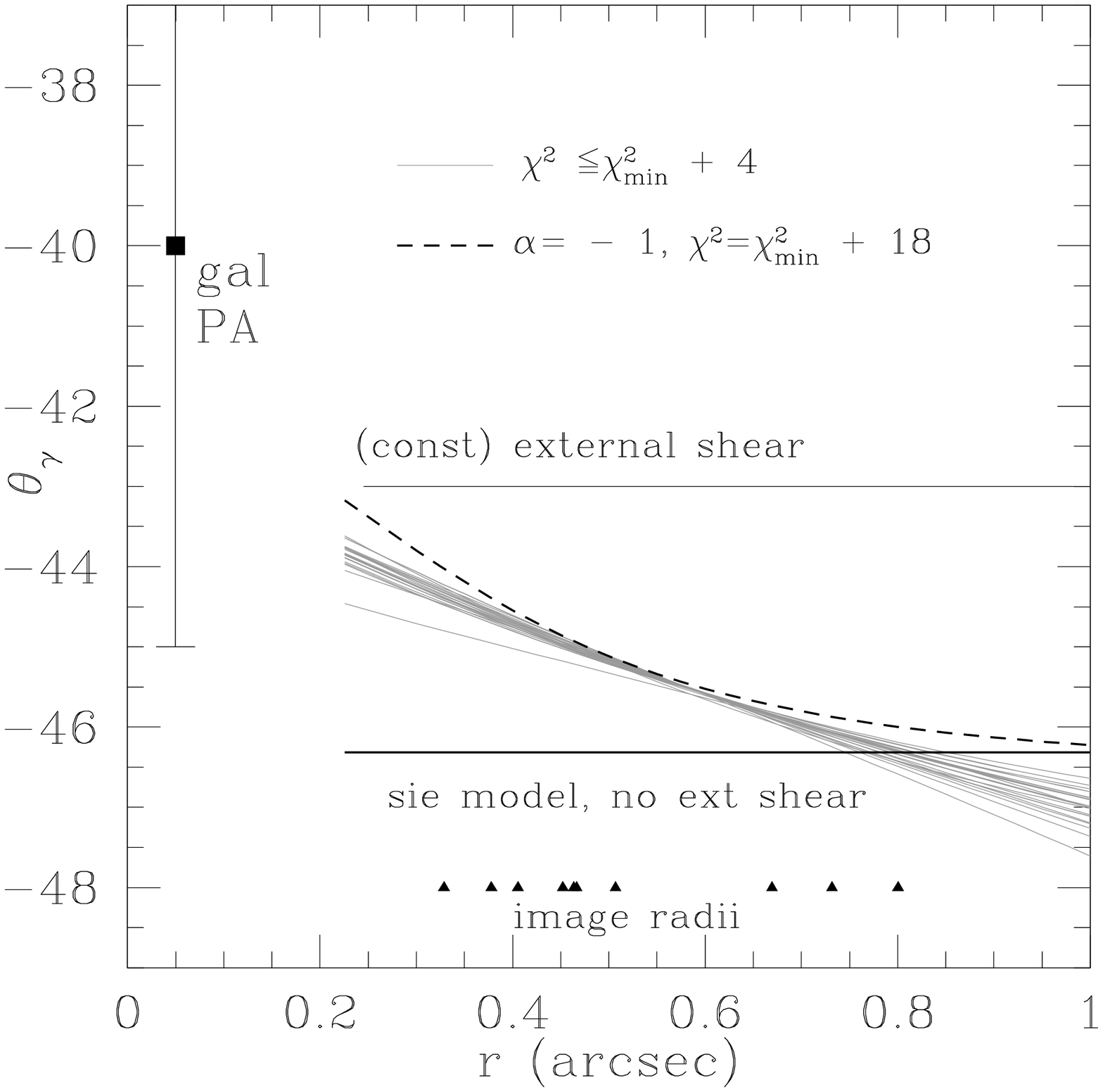}
\end{center}
\caption{The magnitude $\gamma$ (top) and
orientation $\theta_\gamma$ (bottom) of the
quadrupole of the deflection.  Note that here, unlike in the
text, the angle is
given in terms of position angle (i.e. $x = -r \sin \theta, \;
y = r cos \theta$).
A quadrupole of a simple external shear is illustrated
by the horizontal lines in each panel.
A range of acceptable models are shown by the
light gray lines, and the heavy dark line and the dashed line
show two unacceptable models.  We
include some of our hybrid
two-galaxy models to show that the quadrupole structure
is stable when we allow the mass distribution additional degrees of freedom for
its angular structure.  The position angle of the lens galaxy is shown
at its measured value of $-40 \pm 5^\circ$ (Sykes et al. \protect\cite{sykes}).
}
\end{figure}

The second simple physical property of the models we can compare is the
quadrupole of the
lens.  We define the quadrupole by the appropriate multipole moment of the
angular deflection,
\begin{equation}
\left(
   \begin{array}{r}
     \gamma \cos 2 \theta_\gamma\\
      \gamma \sin 2 \theta_\gamma
   \end{array}
   \right)
   = { 1\over \pi r^2 } \int d\theta
\left[{\partial \over \partial \theta}  \phi \right]
     \left(
     \begin{array}{r}
    - \sin 2 \theta\\
     \cos 2 \theta
   \end{array}
      \right),
\en
including both the lens galaxy and
the external shear in the potential.  We have
defined the quadrupole such that the quadrupole of an SIE model is proportional
to $r^{-1}$ and the quadrupole of an external shear is independent of radius.
An external shear $(\gamma/2) \cos(2 \theta - 2 \theta_\gamma)$ produces a
quadrupole moment $\gamma$ at orientation $\theta_\gamma$.  Figure 6
shows the amplitude and major axis PA of the quadrupole moment as a function
of radius for the statistically acceptable models.\footnote{Note that
the PA is related to $(x,y)$ via $x = -r \sin \theta, \;
y = r \cos \theta$, while the angles in the equations obey
the more standard definition $x = r \cos \theta, \;
y = r \sin \theta$.}
The measured PA
of the major axis of the lens,
$-40^\circ\pm5^\circ$ (Sykes et al \cite{sykes}) is shown for comparison.
  All the sucessful models
show a similar decline and flattening of the strength of the quadrupole
combined with a small $ \sim 3^\circ$ swing in orientation. The poorly fitting
power law models also have very different
quadrupole profiles from the acceptable
models.  When we tested the composite models,
where the properties of the monopole
and quadrupole are largely decoupled, the monopole and
quadrupole deflection profiles
of the best two component models were little changed from the acceptable
single profile models. Some
examples are included in Figure 6.

Although B~1933+507 has yet to show the variability needed to measure a time
delay (Biggs et al. \cite{biggs}),
the constraints on the mass distribution are
significantly better than for the lenses
with delay measurements.  Figure 7 shows
the variation in the predicted longest delay
between the compact radio components
(between images 1 and 6) as a function of
the $\chi^2$ statistic for the various
models.  The delay depends on the outer
cutoff of the mass distribution, with the
delay rising as the mass distribution is
truncated at smaller radii and the depth
of the central potential decreases.  This
is a common feature of lens time delays (see
Keeton \& Kochanek \cite{kk}, Impey et al. \cite{impey},
Koopmans \& Fassnacht \cite{koop},
Witt, Mao \& Keeton \cite{wmk00}).  Formally we find a delay range of
$(10.6^{+2.4}_{-1.1})h^{-1}$~days.
As seems to be typical of lens systems (see Keeton \& Kochanek \cite{kk},
Impey et al. \cite{impey}),
the ratios of the delays between the
images depend weakly on the model.  For example,
the ratio of the delay between images 1 and 4
to the delay between images 1 and 6 changes
by only $\sim$ 3\% ($< 0.4h^{-1}$~days) for the range
of the statistically acceptable models.
Models with central cusps which are more realistic than
the $\alpha > 1$ power law models will have
an outer break radius similar to the pseudo-Jaffe models.
As the models are varied by decreasing the break radius the
time delays will rise.
\begin{figure}[h]
\begin{center}
\leavevmode
\epsfxsize=3.5in \epsfbox{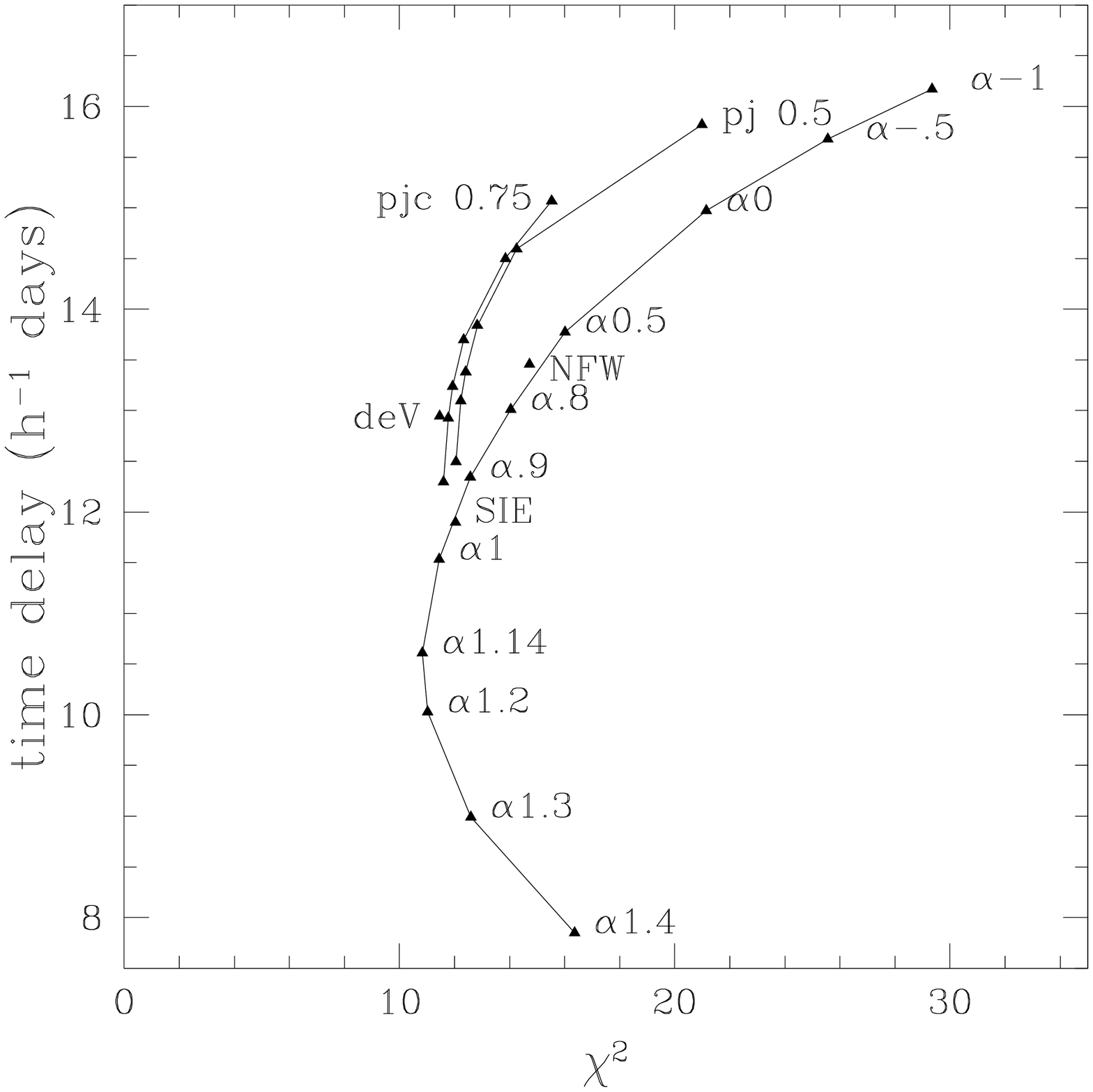}
\end{center}
\caption{The dependence of the longest time
delay (between images 1 and 6) on the model
 and the fit statistics $\chi^2$.
 There is a well defined prediction for the time delay
 including the model uncertainties.  The variation in the time delay shows the
 characteristic pattern that the time delay
rises as the mass distribution becomes more
 centrally concentrated.   Models
with central cusps which are more realistic than the $\alpha >1$
 power law models would have an outer break radius
similar to that of the pseudo-Jaffe
 models.  As the central break radius approached the
Einstein radius of the lens, the
 time delay and (presumably) the $\chi^2$ would rise in a similar manner.
}
\end{figure}

\section{Conclusions and Prospects}

After fitting a wide range of parameterized, ellipsoidal mass distributions
to the B~1933+503 system, we conclude that the mass distribution must have
an approximately flat rotation curve with a cuspy central density distribution.
Over the range spanned by our acceptable models the rotation curve deviates
from the average by less than
$\sim $8.5\%  between the inner and outer images
($1.7h^{-1}$~kpc to $4.1h^{-1}$~kpc).  The central cusp needs to be
quasi-isothermal ($\rho \propto r^{-2}$) and models
with shallow cusps ($\rho \propto 1/r$) fail to fit the data.  Traditional
power law models with finite central core radii are inconsistent
with the data, as we would expect from our current understanding of galaxy
mass distributions.  Both the monopole and quadrupole deflection profiles
of the acceptable models are similar and stable to increasing the complexity
of the models further.  The historically popular power-law models for
gravitational lenses should be abandoned in favor of the lensing equivalents
of the $\eta$-models (Dehnen \cite{dehnen}, Tremaine et al.
\cite{tremaine94}).  These models are
parametrized by a central cusp exponent and break radius between an interior
cusp and a more rapidly declining outer density profile, rather than a core
radius and the slope of the asymptotic density profile.

So far, all lenses which have been analyzed to determine the radial mass
distribution of the lens prefer mass distributions corresponding to
nearly flat rotation curves.  This is true of MG~1654+1346
(Kochanek \cite{csk1654}),
MG~1131+0456 (Chen et al. \cite{ckh95}),
Q~0957+561 (Keeton et al. \cite{keeton2k}),
PG~1115+080 (Kochanek et al. \cite{kkm}) and B~1933+507 (this paper).  No
lens with an image distribution suitable for determining the radial
mass profile has ever found results which are grossly inconsistent with
a nearly flat rotation curve.  Similar results are also found for
observations of nearby early-type galaxies with X-ray halos (Fabbiano
\cite{fabbiano}) and modern stellar dynamical
studies (e.g. Rix et al. \cite{rixetal}).

The predicted time delays of B~1933+503 are relatively well constrained at
$(10.6^{+2.4}_{-1.1})h^{-1}$~days between images 1 and 6 over the wide range of
possible mass distributions we explored. 
A $\sim$22\% modeling uncertainty is
encouraging because there are excellent prospects for further improving
the model constraints. Improved radio maps with a better quantitative
understanding of the mapping between images 2a,~2b,~5 and~7,
astrometry of images 1a~and~8, and tighter constraints on the flux
of any central images would all be useful.  The time delay ratios
between the images can constrain the models if measured to 3\% or better.
Deeper infrared images of the system, particularly with the Hubble Space
Telescope, could enormously improve the relative astrometry between
the lens galaxy and the radio sources and accurately determine the
structure of the lens galaxy (ellipticity, orientation $\cdots$). Most
importantly, the existing HST images strongly suggest the presence of
lensed images of the radio source host galaxy.  Deep imaging to measure
the shape of the Einstein ring image of the host would dramatically improve
the constraints (see Kochanek, Keeton \& McLeod \cite{kkm}).

\bigskip
\noindent Acknowledgements:
CSK was supported by NASA Astrophysics 
Theory Program grants NAG5-8831 and NAG5-9265.  
Support for the CASTLES project was provided by NASA through grant numbers
GO-7495,  GO-7887, and GO-8175 from the Space 
Telescope Science Institute, which is
operated by the Association of Universities for Research in Astronomy, Inc. 
JDC thanks M. White for conversations and the ITP for
hospitality.  JDC was supported in part by the
National Science Foundation under Grant No. AST-0074728 and
while at the ITP under Grant No. PHY94-07194.
We thank the referee for helpful suggestions and J. Munoz for
discussions on the input constraints and comments on the draft.

\bigskip

\appendix
\section{Summary of Models}

Here we present the analytic expressions for the model surface densities
in terms of the ellipsoidal coordinate $m$, defined as
$m^2 = x'^2+y'^2/q^2$,
where $x'$ and $y'$ are Cartesian
coordinates aligned with the major and minor axes of the lens.
The singular isothermal ellipsoid (SIE) is
defined by
\begin{equation}
  \kappa_{SIE}(m) = { 1 \over 2 } { b \over |m|} \; \; \; {\rm SIE},
\end{equation}
and the power-law models are defined by
\eq
 \kappa_\alpha (m) = { 1 \over 2 } b^{2-\alpha}
\left(s^2 + m^2\right)^{\alpha/2-1}  \; \; \; {\rm power \; law} \; .
\en
The power-law model with $\alpha=1$ and $s=0$ is the SIE model.  The
pseudo-Jaffe model is the difference of two softened isothermal
ellipsoids ($\alpha=1$
with a finite core radius), where
\begin{equation}
  \kappa_{PJ}(m) = { 1 \over 2 } b \left[ \left(s^2+m^2\right)^{-1/2}
         - \left(a^2+m^2\right)^{-1/2} \right]
 \; \; \; {\rm pseudo-Jaffe}
\end{equation}
with $ s \leq a $.  The core radius of the profile is $s$ and the break radius
is $a$.
It matches the SIE model for $s\rightarrow 0$ and $a \rightarrow \infty$,
the general $\alpha=1$ power-law model for $a\rightarrow \infty$ and
the $\alpha=-1$ power
law model in the limit that $s \rightarrow a$ with $b(a-s)$ kept constant.
The de Vaucouleurs model is defined by
\begin{equation}
    \kappa_{dV}(m) = \frac{b}{2 {\cal N} R_e} \exp\left[-k \left( \frac{m}{R_e}
\right)^{1/4}
\right]  \; \; \; {\rm de Vaucouleurs}
\end{equation}
where
\eq
k = 7.67 \; {\rm and} \;
{\cal N} = \int_0^\infty v e^{-kv^{1/4}} \, dv = 1.683 \times 10^{-3}
\en
and the NFW model by
\begin{equation}
   \kappa_{NFW} (m) =  2 \kappa_s { 1- {\cal F}(m/a) \over (m/a)^2 - 1 }
 \; \; \; {\rm NFW}
\end{equation}
where
\begin{equation}
{\cal F}(m/a) = \left\{
\begin{array}{ll}
\frac{1}{\sqrt{m^2/a^2 - 1}} \tan ^{-1} \sqrt{m^2/a^2 - 1}
& \; \; m/a > 1 \\
\frac{1}{\sqrt{1- m^2/a^2}} \tanh ^{-1} \sqrt{1-m^2/a^2}
 & \; \; m/a < 1 \\
1 & m/a = 1
\end{array} \right.
\end{equation}

\end{document}